\documentclass[aps,prb,twocolumn,superscriptaddress,showpacs]{revtex4-1}
\usepackage{graphicx,amsmath,color,placeins}
\usepackage{ulem}

\begin{document}
	\title{Band Geometry Induced High-Angular Momentum Excitonic Superfluid in Gapped Chiral Fermion Systems}
	
	\author{Huaiyuan Yang}
	\affiliation{State Key Laboratory for Artificial Microstructure and Mesoscopic Physics, Frontier Science Center for Nano-optoelectronics and School of Physics, Peking University, Beijing 100871, P. R. China}
	\author{Yuelin Shao}
	\email{ylshao@iphy.ac.cn}
	\affiliation{Beijing National Laboratory for Condensed Matter Physics and Institute of Physics, Chinese Academy of Sciences, Beijing 100190, P. R. China}
	\affiliation{School of Physical Sciences, University of Chinese Academy of Sciences, Beijing 100049, P. R. China}
	\author{Xi Dai}
	\email{daix@ust.hk}
	\affiliation{Department of Physics, Hong Kong University of Science and Technology, Clear Water Bay, Kowloon 999077, Hong Kong}
	\author{Xin-Zheng Li}
	\email{xzli@pku.edu.cn}
	\affiliation{State Key Laboratory for Artificial Microstructure and Mesoscopic Physics, Frontier Science Center for Nano-optoelectronics and School of Physics, Peking University, Beijing 100871, P. R. China}
	\affiliation{Interdisciplinary Institute of Light-Element Quantum Materials, Research Center for Light-Element Advanced Materials, and Collaborative Innovation Center of Quantum Matter, Peking University, Beijing 100871, People's Republic of China}
	\affiliation{Peking University Yangtze Delta Institute of Optoelectronics, Nantong, Jiangsu 226010, People's Republic of China}
	\date{\today}

	\begin{abstract}
		We study the exciton condensation in the heterostructures where the electron layer and hole layer formed by gapped chiral Fermion (GCF) systems are separately gated.
		High-angular momentum such as p- and d-wave like excitonic pairing may emerge when the gap of the GCF systems is small compared to the Fermi energy, and the chiral winding number of the electrons and holes are the same.
		This is a result of the non-trivial band geometry and can be linked to the Berry curvature when projected onto the Fermi surface.
		In realistic systems, we propose that staggered graphene and magnetic topological surface states are promising candidates for realizing p-wave exciton superfluid, and anomalous Hall conductivity can be used as a signature in experiments.

	\end{abstract}
	\maketitle
	\clearpage
	\textbf{Introduction}
	
	Composed of a bound electron-hole pair, an exciton is a charge-neutral Bosonic quasiparticle.
	In some specific systems such as intrinsic excitonic insulators or heterostructures formed by separated electron-hole layers, excitons can undergo Bose-Einstein condensation and turn into superfluid \cite{keldysh1965possible, kohn1967excitonic, jerome1967excitonic, RevModPhys.40.755}.
	Numerous theoretical works have predicted many non-trivial phenomena in the p-wave exciton superfluid, including topologically protected vortex zero modes, charge fractionalization, parity anomaly and non-Abelian braiding \cite{PhysRevLett.101.246404, PhysRevLett.103.066402, wang2019prediction, PhysRevLett.128.106804}.
	However, all experimentally realized exciton condensation systems are s-wave like, including graphene and transition metal dichalcogenides (TMD) systems \cite{ponomarenko2011tunable, liu2017quantum, li2017excitonic, jia2022evidence, sun2022evidence, gu2022dipolar, wang2019evidence, ma2021strongly, chen2022excitonic}.
	This has strongly hindered the progress in the field of this novel physics.
	Many proposals have been made in recent years to achieve p-wave excitonic pairing.
	Some discussed tunneling assisted p-wave excitonic insulator \cite{PhysRevLett.112.176403, zhu2019gate, PhysRevLett.131.046402}, but it is not a superfluid state as the phase transition there only corresponds to the breaking of discrete symmetries \cite{PhysRevLett.124.197601}.
	Perfetto and Stefanucci found that floquet engineering can help to realize nonequilibrium p-wave excitonic insulators \cite{PhysRevLett.125.106401}.
	However, an achievable system for equilibrium p-wave exciton superfluid is still absent.
	One feature in common for these previous works is that the band geometry effects on exciton condensation have been neglected, and the band edges are described by the effective mass model.
	In the recent decade, however, many works have shown that this approximation breaks down when considering the exciton energy spectrum \cite{PhysRevLett.115.166802, PhysRevLett.115.166803} and optical selection rules \cite{PhysRevLett.120.087402, PhysRevLett.120.077401}.
	One can expect that the band geometry may also influence exciton condensation.
	This is especially true in gapped chiral Fermion (GCF) systems, where the band geometry matters in many physical properties \cite{PhysRevLett.115.166802, PhysRevLett.115.166803, PhysRevLett.120.087402, PhysRevLett.120.077401, PhysRevB.95.035311, PhysRevLett.110.197402, PhysRevLett.108.196802, PhysRevLett.110.016806, PhysRevLett.112.166601, PhysRevLett.113.156603}.
	Many two-dimensional (2D) materials can be described by GCF, e.g. gapped graphene monolayer and multilayers\cite{PhysRevLett.96.086805}, gapped topological surface states\cite{PhysRevB.84.045403}, and many TMD monolayers like MoS$_2$ \cite{PhysRevLett.108.196802}.
	Therefore, a detailed exploration of the exciton condensation in GCF systems with band geometry taken into account is highly desired.
	In this letter, we study the band geometry effects on exciton condensation in the charge-neutral hybrid heterostructures.
	The BCS regime is considered, where the separated electron and hole layers consisting of GCFs have definite Fermi surfaces (FSs).
	Through the analysis of the gap equation and numerical calculations, we find that high-angular momentum such as p-wave and d-wave like exciton condensation may appear when the chiral winding number of the electron and hole GCF systems are both 1 or both 2, and the gap of the GCFs is small compared to the gating voltage.
    Further, under FS projection approximation, the angular momentum of the exciton condensation is found directly related to the Berry curvature flux enclosed in the electron and hole FSs.
	By increasing the total Berry curvature flux enclosed in the FSs, for example, increasing the charge density per layer, the exciton condensation might experience a s- to p-wave  transition which will induce an integer jump of the anomalous Hall conductivity.
	For	the material realization of the p-wave exciton condensation, we propose that staggered graphene and magnetic topological surface states are promising candidates.

	\textbf{Results}

	\textit{Model-}We first introduce the 2D isotropic GCF model Hamiltonian for a single valley\cite{PhysRevLett.120.077401}:
	\begin{equation}\label{gcf}
		H_{\text{GCF}}=\left(\begin{array}{cc}
			D & \alpha k^{|w|} e^{i w \theta_{\boldsymbol{k}}} \\
			\alpha k^{|w|} e^{-i w \theta_{\boldsymbol{k}}} & -D
		\end{array}\right).
	\end{equation}
	Here $2D>0$ is the energy gap at ${\boldsymbol{k}}=0$ measured from the valley and $\theta_{\boldsymbol{k}}$ is the polar angle of  ${\boldsymbol{k}}$.
	$|w|=1$ and $|w|=2$ are the most common cases.
	TMDs and gapped graphene correspond to $|w|=1$, while gated bilayer graphene\cite{PhysRevLett.96.086805} and topological excitonic insulator AsO and Mxene monolayers \cite{yang2023spintriplet, doi:10.1021/acs.nanolett.6b03118, doi:10.1063/1.4983781} correspond to $|w|=2$.
	The eigenenergies are $\varepsilon_{c / v}= \pm \epsilon_k= \pm \sqrt{D^2+\alpha^2 k^{2|w|}}$, with eigenstates:
	\begin{equation}
		|c \boldsymbol{k}\rangle=\left(\begin{array}{c}
			\cos \frac{\phi_k}{2} \\
			\sin \frac{\phi_k}{2} e^{-i w \theta_{\boldsymbol{k}}}
		\end{array}\right), \quad|v \boldsymbol{k}\rangle=\left(\begin{array}{c}
			\sin \frac{\phi_k}{2} e^{i w \theta_{\boldsymbol{k}}} \\
			-\cos \frac{\phi_k}{2}
		\end{array}\right),
	\end{equation}
	where $\phi_{k}=\arccos \left(D / \epsilon_k\right), \phi_{k} \in(0, \pi/2)$.
	Using this gauge, the lowest energy exciton state is a 1s state in the 2D hydrogen model \cite{PhysRevLett.115.166803, PhysRevLett.120.077401}.
	Incorrect conclusions may be drawn if the gauge is not reasonably chosen~\cite{PhysRevLett.111.086804}.

	\begin{figure}[h]
		\includegraphics[width=1.0\linewidth]{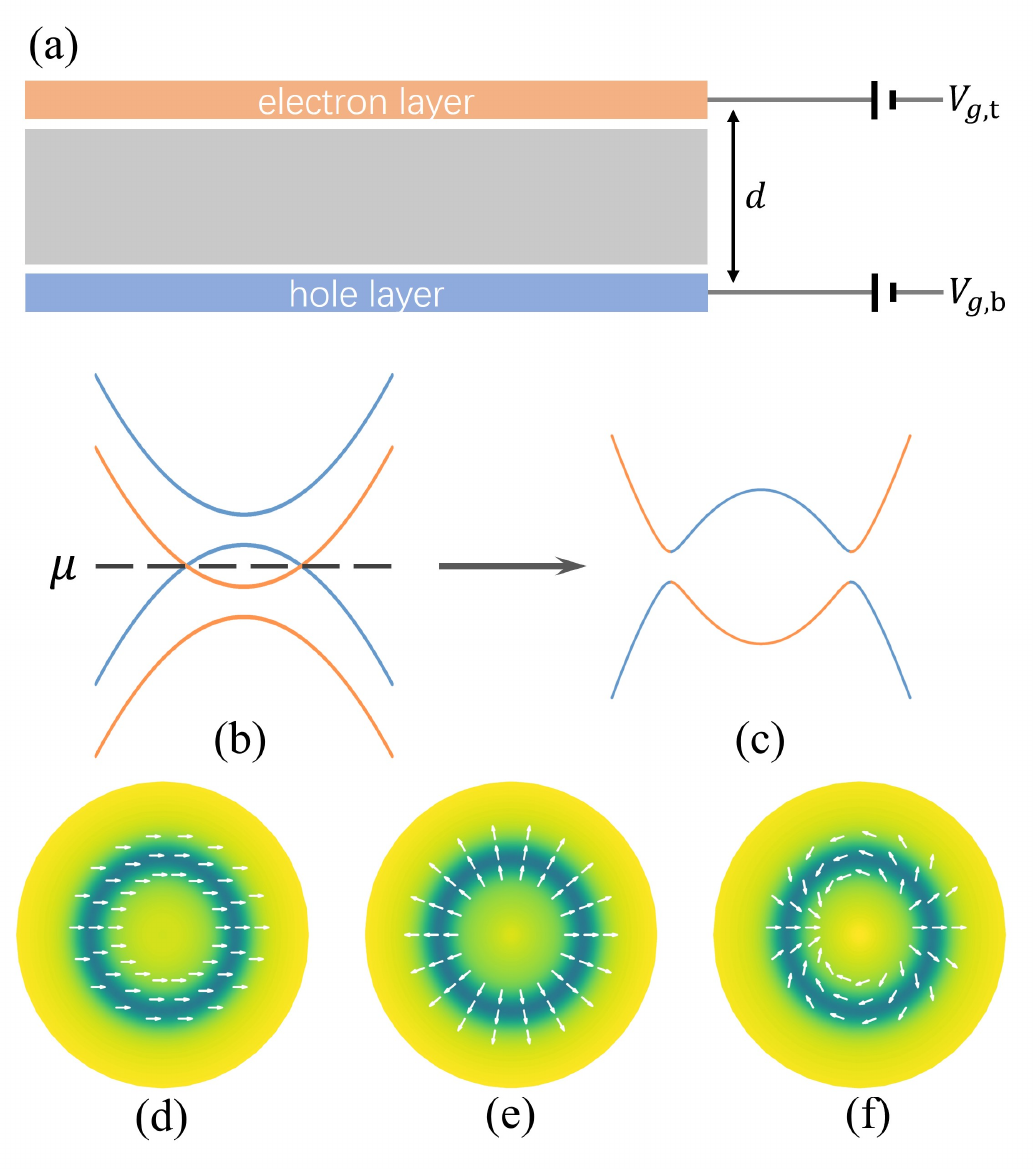}
		\caption{ \label{fig1}
			(a) Schematic for the exciton condensation heterostructure.
			The electron and hole layers are separated in the $z$ direction with a width $d$.
			(b) At charge neutrality point, the electron and hole FS are perfectly nested.
			(c) The excitonic instability is favoured, and a gap opens.
			Here the orange (blue) lines indicate the bands from the top (bottom) layer.
			(d, e, f) s-, p-, d-wave like excitonic order parameter.
			The direction of the arrow denotes the phase angle and the darkness of the colormap indicates the magnitude.
		}
	\end{figure}

	The double-layer exciton condensation heterostructure is demonstrated in Fig.1(a).
	Each layer is described by a GCF model.
	The top (bottom) layer are separately gated by $V_\text{t}$ ($V_\text{b}$) so that the top (bottom) layer has electron (hole) FS.
	In this letter, we assume the charge neutral case ($V_{g,\text{t}}=-V_{g,\text{b}}=-V_g$), where the electron and hole FSs are perfectly nested as shown in Fig.1(b), favoring the excitonic instability the most.
	After taking the electron-hole interaction into consideration, the formation of inter-layer exciton condensation will open a gap (Fig.1(c)), and the system evolves from a semimetal to an excitonic insulator.
	The non-interacting Hamiltonian for the two-layer system in the two-band basis reads:
	\begin{equation}\label{h0}
		H_0=\sum_{\boldsymbol{k}} \Psi_{\boldsymbol{k}}^{\dagger}\left(\begin{array}{cc}
			V_{g,\text{t}} + \varepsilon_{c\boldsymbol{k}, \text{t}} & 0 \\
			0 & V_{g,\text{b}} + \varepsilon_{v\boldsymbol{k}, \text{b}}
		\end{array}\right) \Psi_{\boldsymbol{k}},
	\end{equation}
	where $\Psi_{\boldsymbol{k}}^{\dagger}=\left[c_{c\boldsymbol{k}, \text{t}}^{\dagger}, c_{v\boldsymbol{k}, \text{b}}^{\dagger}\right]$.
	Here $c_{c\boldsymbol{k}, \text{t}}$ ($c_{v\boldsymbol{k}, \text{b}}$) refers to the electrons in the conduction (valence) band of the top (bottom) layer in Fig.1b.
	The interlayer interaction between these two bands reads:
	\begin{equation}
		\begin{aligned}
			H_{\text{int}}= &\sum_{\boldsymbol{k}_1, \boldsymbol{k}_2, \boldsymbol{q}} V_{\text{inter}}(q) \Lambda_{cc}\left(\boldsymbol{k}_1, \boldsymbol{k}_1-\boldsymbol{q}\right) \Lambda_{vv}\left(\boldsymbol{k}_2, \boldsymbol{k}_2+\boldsymbol{q}\right) \\
			& \times c_{c\boldsymbol{k}_1, \text{t}}^{\dagger} c_{v\boldsymbol{k}_2, \text{b}}^{\dagger} c_{v \boldsymbol{k}_2+\boldsymbol{q}, \text{b}} c_{c \boldsymbol{k}_1-\boldsymbol{q}, \text{t}},
		\end{aligned}
	\end{equation}
	where $V_{\text{inter}}(q)$ is the interlayer screened Coulomb interaction,
	$\Lambda_{ab}\left(\boldsymbol{k}, \boldsymbol{p}\right) \equiv\left\langle a\boldsymbol{k} \mid b\boldsymbol{p}\right\rangle$ is the form factor where $a, b \in \{c, v\}$.
	In GCF model, it deviates from unity:
	\begin{equation} \label{form_factor}
		\begin{aligned}
			\Lambda_{cc} \left(\boldsymbol{k}, \boldsymbol{p}\right) = & \cos \left(\frac{\phi_{c k}}{2}\right) \cos \left(\frac{\phi_{c p}}{2}\right)+ \\
			& \sin \left(\frac{\phi_{c k}}{2}\right) \sin \left(\frac{\phi_{c p}}{2}\right) e^{i m\left(\theta_{\boldsymbol{k}}-\theta_{\boldsymbol{p}}\right)}, \\
			\Lambda_{vv} \left(\boldsymbol{k}, \boldsymbol{p}\right) = & \cos \left(\frac{\phi_{v k}}{2}\right) \cos \left(\frac{\phi_{v p}}{2}\right)+ \\
			& \sin \left(\frac{\phi_{v k}}{2}\right) \sin \left(\frac{\phi_{v p}}{2}\right) e^{-i n\left(\theta_{\boldsymbol{k}}-\theta_{\boldsymbol{p}}\right)},
		\end{aligned}
	\end{equation}
	where $m$ ($n$) is the chiral winding number for the conduction (valence) band.
	Later we will see that this band geometry effect plays an essential role in the excitonic pairing.
	\textit{Gap Equation-}Using standard Hartree-Fock approximation, we obtain the gap equation for the interlayer excitonic pairing at 0K:
	\begin{equation}\label{gap_eq}
		\Delta_{\boldsymbol{k}}= \Sigma_{\boldsymbol{p}} V_{\text{inter}}(|\boldsymbol{k}-\boldsymbol{p}|) \Lambda_{cc} \left(\boldsymbol{k}, \boldsymbol{p}\right) \Lambda_{vv} \left(\boldsymbol{p}, \boldsymbol{k}\right)  \frac{\Delta_{\boldsymbol{p}}}{2E_{\boldsymbol{p}}},
	\end{equation}
	where $2E_{\boldsymbol{p}}$ is the renormalized energy gap.
%\begin{equation}\label{Elabel}
%E_{\boldsymbol{p}} = \sqrt{(V_\text{t} + \varepsilon_{c\boldsymbol{k}, \text{t}}- V_\text{b} - \varepsilon_{v\boldsymbol{k}, %\text{b}})^2/4+\Delta_{\boldsymbol{p}}^2 },
%\end{equation}
%and $f_{\boldsymbol{p}}$ is the occupation number difference between the lower and upper renormalized band.
	%
	For brevity, we leave the details of the full Hartree-Fock equations in Methods.
	To analyze the excitonic pairing characteristics, we decompose the gap function into different angular momentum channels: $\Delta_{\boldsymbol{k}}=\Sigma_J \left|\Delta_k\right|_J e^{i J \theta_{\boldsymbol{k}}}, J=0, \pm 1, \pm 2 \ldots$, and $\left|\Delta_k\right|_J$ is independent of the polar angle \cite{PhysRevLett.111.086804}.
	In this way, Eq. \ref{gap_eq} can be rewritten as independent equations for different $J$s:
	\begin{equation}\label{J_gap_eq}
		\begin{aligned}
			\left|\Delta_k\right|_J= & \Sigma_{\boldsymbol{p}} e^{-i J\left(\theta_{\boldsymbol{k}}-\theta_{\boldsymbol{p}}\right)} V_{\text{inter}}(|\boldsymbol{k}-\boldsymbol{p}|) \\
			& \Lambda_{cc} \left(\boldsymbol{k}, \boldsymbol{p}\right) \Lambda_{vv} \left(\boldsymbol{p}, \boldsymbol{k}\right)  \frac{\left|\Delta_p\right|_J }{2E_{p}}.
		\end{aligned}
	\end{equation}
	After inserting Eq. \ref{form_factor} into Eq. \ref{J_gap_eq},  and seeing that the imaginary parts of the exponentials are odd and vanish after the angle integration, we get:
	\begin{widetext}
		\begin{equation}\label{detail}
			\begin{aligned}
				\left|\Delta_k\right|_J= & \Sigma_{\boldsymbol{p}} V_{\text{inter}}(|\boldsymbol{k}-\boldsymbol{p}|) \frac{\left|\Delta_p\right|_J}{2E_{p}}f_{p}  \times [ \cos \left(\frac{\phi_{c k}}{2}\right) \cos \left(\frac{\phi_{c p}}{2}\right) \cos \left(\frac{\phi_{v k}}{2}\right) \cos \left(\frac{\phi_{v p}}{2}\right) \cos J(\theta_{\boldsymbol{k}}-\theta_{\boldsymbol{p}})   \\
				& +\cos \left(\frac{\phi_{c k}}{2}\right) \cos \left(\frac{\phi_{c p}}{2}\right) \sin \left(\frac{\phi_{v k}}{2}\right) \sin \left(\frac{\phi_{v p}}{2}\right) \cos (J-n)(\theta_{\boldsymbol{k}}-\theta_{\boldsymbol{p}}) \\
				& +\sin \left(\frac{\phi_{c k}}{2}\right) \sin \left(\frac{\phi_{c p}}{2}\right) \cos \left(\frac{\phi_{v k}}{2}\right) \cos \left(\frac{\phi_{v p}}{2}\right) \cos (J-m)(\theta_{\boldsymbol{k}}-\theta_{\boldsymbol{p}}) \\
				& +\sin \left(\frac{\phi_{c k}}{2}\right) \sin \left(\frac{\phi_{c p}}{2}\right) \sin \left(\frac{\phi_{v k}}{2}\right) \sin \left(\frac{\phi_{v p}}{2}\right) \cos (J-m-n)(\theta_{\boldsymbol{k}}-\theta_{\boldsymbol{p}}) ].
			\end{aligned}
		\end{equation}
	\end{widetext}
	In a simplified model when the interlayer Coulomb interaction $V_{\text{inter}}(k)$ is a constant independent of $\boldsymbol{k}$, i.e., it is a short-range interaction,
	one can perform the angle integration, and realize immediately that only when $J=0, n, m, m+n$, can $\left|\Delta_k\right|_J$ be non-zero.
	As $\phi \in(0, \pi/2)$, $\cos(\phi/2)>\sin(\phi/2)$, the pairing potential in $J=0$ channel is the strongest, and the excitonic pairing is s-wave like, as shown in Fig.1(d).

	\begin{figure}[h]
		\includegraphics[width=1.0\linewidth]{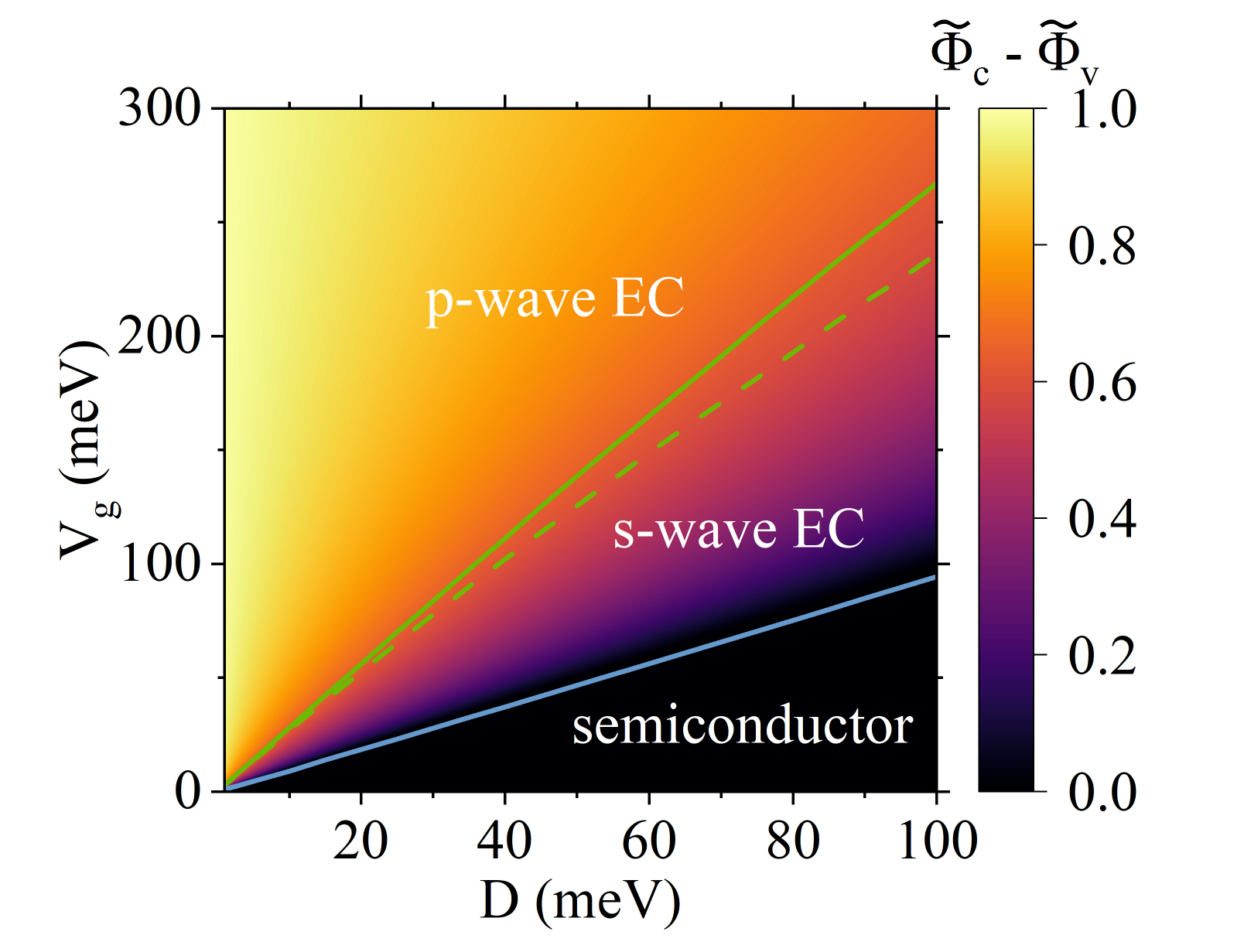}
		\caption{ \label{fig2}
			The phase diagram when the electron and hole layers are both $m=1$ GCF model, and $\alpha=10^6 m/s$
			The solid lines are obtained by self-consistent calculations, and the dashed line is determined by the FS projection.
			The colorbar indicates the Berry curvature flux.
			EC means exciton condensation.
		}
	\end{figure}

	In realistic systems, $V_{\text{inter}}(k)$ is a long-range interaction, having the form of $V_{\text{inter}}(q) = \frac{2 \pi }{S \varepsilon_r q} e^{-qd}$.
	Here $S$ is the system area, $\varepsilon_r=5$ is the dielectric constant of the heterostructure, and $d=4$ nm is the distance between the two layers.
	We consider two limit cases.
	First, when $D$ of the GCF model is much larger than $\alpha k_F$, then $\phi$ is close to 0, and $\cos(\phi/2) \gg \sin(\phi/2)$.
	$J=0$ channel has the largest pairing strength and the exciton is s-wave like.
	Second, in the opposite limit, when $D \rightarrow 0$, $\cos(\phi/2) \approx \sin(\phi/2) \approx \sqrt2 / 2$, interesting physics occurs.
	When $m=n$, the pairing in $J=m$ channel becomes larger than the $J=0$ and $J=2m$ channels as the second and third terms within the brackets of Eq.~\ref{detail} merge.
	The excitonic pairing becomes p-wave like when $m=1$ or d-wave like when $m=2$, as shown in Fig.1(e, f).
	Assuming that the electron and hole layers are constituted of the same material, we obtain the numerical results of different pairing order parameters using Hartree-Fock equations, and plot the phase diagram with different $D$s and $V_g$s for $m=n=1$ in Fig.2.
	The excitonic pairing is s-wave like when $D$ is large, e.g. the TMD systems where $D\approx1$ eV.
	For systems with small $D$, the long-sought high-angular momentum exciton condensation may appear, and its material realization will be discussed later.
	\textit{FS Projection-}The correlation between the angular momentum of the exciton condensation and the band geometry can be seen more clearly when we project the interaction onto the FS.
	The projection is reasonable in the BCS regime as the pairing order parameter is mainly distributed around the narrow region of the FS \cite{PhysRevLett.74.1633, littlewood2004models, PhysRevB.74.165107, PhysRevB.78.121401, PhysRevB.79.235329, PhysRevLett.111.086804, PhysRevLett.110.146803, PhysRevLett.119.257002}, which is also shown in Figs.~1(d-f).
	Therefore, we can use the FS projected interlayer interaction, defined as $U_{m, n}\left(k_F, \theta_1, \theta_2\right) \equiv V_{\text{inter}}\left(k_F, \theta_1, \theta_2\right) \Lambda_{c c}\left(k_F, \theta_1, \theta_2\right) \Lambda_{v v}\left(k_F, \theta_2, \theta_1\right)$, to perform an analysis of the
	angular-dependence of the pairing.
	Decomposing $U_{m, n}$, $V_{\text{inter}}$, $\Lambda_{c c}$, and $\Lambda_{v v}$ into different angular momentum channels and using the convolution rules, we obtain
	the interaction in each channel $J$:
	\begin{equation}
		U_{m, n}^J\left(k_F\right)=\sum_{J_1+J_2-J_3=J} V_{\text{inter}}^{J_1}\left(k_F\right) \Lambda_{c c}^{J_2}\left(k_F\right) \Lambda_{v v}^{J_3}\left(k_F\right).
	\end{equation}
	The excitonic pairing is characterized by the channel with the largest $U_{\text{inter}}^J\left(k_F\right)$.
	As shown in Supplementary Information (SI), $\Lambda^J$s satisfy the below relations in general rotational symmetric models, as $\sum_J \Lambda^J=1, \sum_J J \Lambda^J=\frac{\Phi}{2 \pi}$, where $\Phi$ is the Berry curvature flux through the FS.
	In the GCF model, only $\Lambda_{cc}^0$, $\Lambda_{vv}^0$, $\Lambda_{cc}^m$ and $\Lambda_{vv}^{-n}$ are non-zero according to Eq. \ref{form_factor}.
	Then the above relations reduces to $\Lambda_{cc}^0=1- \tilde{\Phi}_c/m, \Lambda_{vv}^0=1-\tilde{\Phi}_v/n, \Lambda_{cc}^{m}= \tilde{\Phi}_c/m, \Lambda_{vv}^{-n}= \tilde{\Phi}_v/n$, where $\tilde{\Phi} = \Phi / 2\pi \in (-|m|/2, |m|/2)$.
	Taking the long-range form of the $V_{\text{inter}}$ into consideration, we obtain the s-to-p transition line for $m=n=1$ in Fig.2 (dashed green line), which is close to the full numerical calculations.
	The excitonic paring is s-wave (p-wave) like when the Berry curvature flux is small (large), which is discussed in detail in SI.
	Fig. 2 also shows two ways to achieve p-wave EC in experiments, decrease $D$ or increase $V_g$, which reflects two approaches to increase the Berry flux, increase the Berry flux concentration or enlarge the FSs.
	It should be noted that in the trivial case where the form factor is unity, the excitonic pairing always takes place in the $J=0$ channel, as $V_{\text{inter}}^{J=0}$ is always the largest for the realistic long-range Coulomb interactions.
	In this way, the GCF model clearly demonstrates the essential role played by band geometry effects in the high-angular momentum excitonic pairing.

	\begin{figure}[h]
		\includegraphics[width=1.0\linewidth]{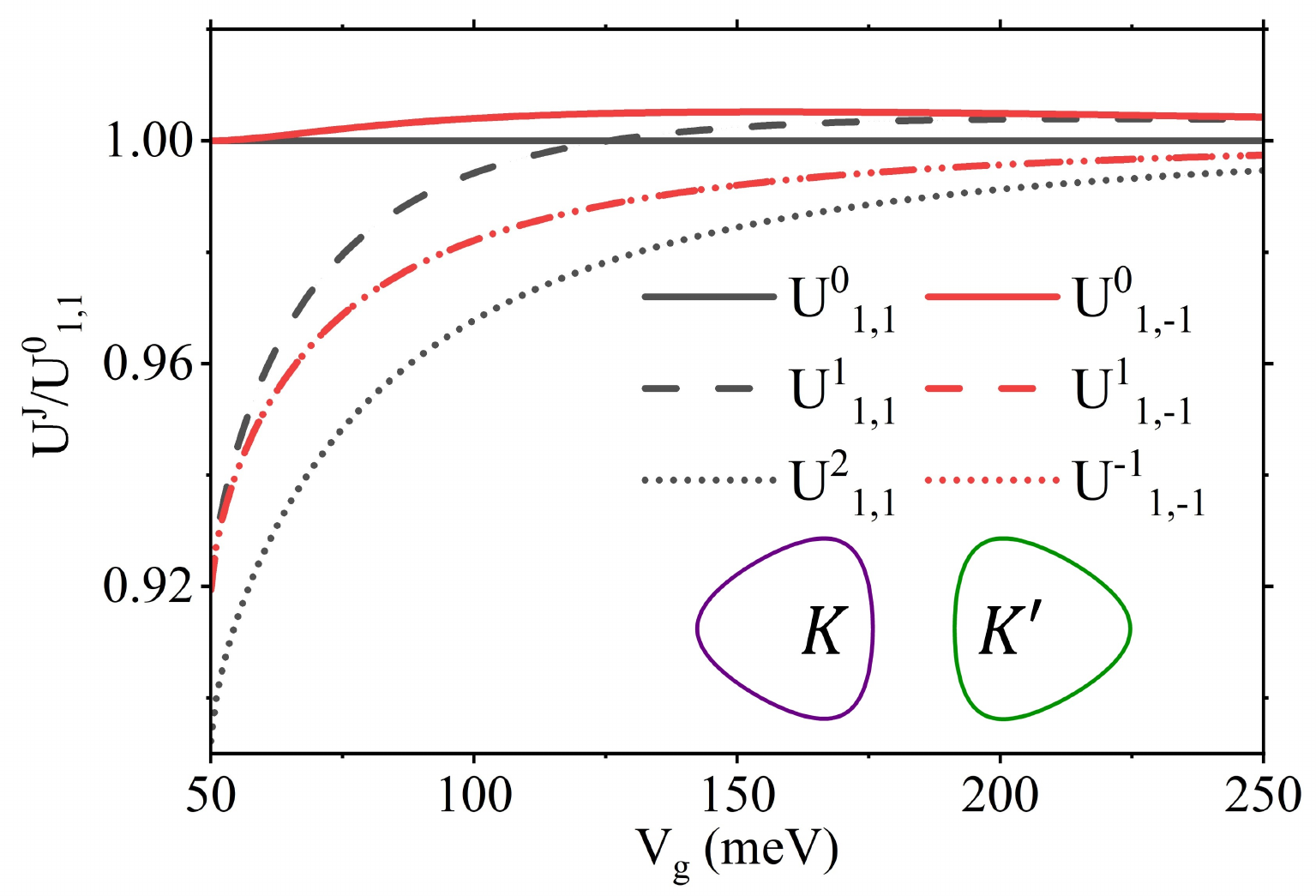}
		\caption{ \label{fig3}
			FS projected interlayer electron-hole interaction $U^J(k_F)$ for $|m|=1$ case, where $k_F$s vary with different $V_g$s.
			Here $D=50$meV.
			Inset: FSs for $K$ and $K^{\prime}$ valleys under finite trigonal warping effect.
		}
	\end{figure}

	\textit{Material Realizaiton-}Now we consider how to achieve the p-wave excitonic condensation in realistic materials (the d-wave realization is discussed in SI).
	In many realistic systems, there are two inequivalent valleys $K$ and $K^{\prime}$ and two spin indices, therefore the intervalley and interspin pairing channels should be considered in addition.
	We consider two general time-reversal symmetric Hamiltonians that correspond to many real 2D materials.
	The first one is $H_1 = v_F (\tau_z \sigma_x k_x - \sigma_y k_y) + \lambda \sigma_z $, where $\tau_i$s and $\sigma_i$s are Pauli matrices in valley and sublattice space, and $\lambda$ is a real positive number.
	This is the case of staggered graphene\cite{jung2015origin, zhou2007substrate, hunt2013massive}.
	In the presence of time-reversal symmetry (TRS), the Berry curvature of the two valleys are opposite, and the chiral winding number is +1 (-1) for the $K$ ($K^{\prime}$) valley.
	The intra- and intervalley electron-hole interlayer interaction $U_{1, 1}^{J}$ and $U_{1, -1}^{J}$ in different $J$ channels are shown in Fig.3, and the intervalley s-wave pairing prevails over the others as $U_{1, -1}^{0}$ is always the largest.
	The second one is $H_2 = v_F (\tau_z \sigma_x k_x - \sigma_y k_y) + \lambda \tau_z \sigma_z s_z $, where $s_z$ is the third Pauli matrix in spin space\cite{PhysRevLett.95.226801} and the last term describes the spin-orbital coupling.
	This corresponds to the low-energy physics of silicene, germanium \cite{PhysRevLett.107.076802} and stanene\cite{PhysRevLett.111.136804}.
	Now the chiral winding numbers are 1, -1, -1, 1 for $K\uparrow$, $K\downarrow$, $K^{\prime} \uparrow$, $K^{\prime} \downarrow$ species.
	The interlayer electron-hole interaciton for the intraspin intravalley, intraspin intervalley, interspin intravalley and interspin intervalley channels are $U_{1, 1}^{J}$, $U_{1, -1}^{J}$, $U_{1, -1}^{J}$, $U_{1, 1}^{J}$ respectively, and the dominating channels are degenerate intraspin intervalley and interspin intravalley s-wave pairings.
    In hexagonal 2D materials, the rotational symmetry of the GCF model is generally degraded down to $C_3$ by the trigonal warping effect.
    %which reads $H_{\text{TW}} = \kappa \left(\begin{array}{cc} 0 & (k_x -  i \tau k_y)^2 \\ (k_x + i \tau  k_y)^2 & 0 \end{array}\right)$, where $\tau=1 (-1)$ for $K$ ($K^{\prime}$) valley.
    %
    As shown in the inset of Fig. 3, the perfect nesting between the intervalley FSs will be broken, and the intravalley pairings will prevail over the intervalley pairings.
    Specifically for the $H_1$ case, Fig. 3 shows that when $V_g$ is large, $U_{1, 1}^{1}$ and $U_{1, -1}^{0}$ are almost the same, and the intravalley p-wave and intervalley s-wave pairings are nearly degenerate.
    Now a finite trigonal warping term will make the intravalley p-wave pairing channel the dominant one.
    Therefore, staggered graphene is a candidate material for the electron and hole layers in Fig. 1(a) to achieve the p-wave exciton condensation.
	Another way to realize the p-wave exciton condensation is to consider the TRS-breaking one-valley systems, such as surface states of the magnetic topological insulator \cite{PhysRevLett.98.106803, PhysRevLett.103.066402}.
	For a thin film of the topological insulator, the surface state Hamiltonian is $H_{\text{t/b}} = \pm v_F (k_x s_y - k_y s_x)$ for the top and the bottom surface, and they can be used as the electron and hole layer respectively under gating in the heterostructure Fig.1(a).
	When the topological insulator becomes magnetic, by either intrinsic or proximate magnetism in the $z$ direction, a $M s_z$ term appears and opens a gap on both surfaces.
	Their chiral winding number are both +1, and the electron-hole interlayer interaction is $U_{1, 1}^{J}$ in different $J$ channels assuming $M<0$.
	According to Fig.2, the excitonic pairing becomes p-wave when the gating $V_g$ is larger than a moderate critical value, as long as $M$ is not too large.
	Therefore, the magnetic topological insulator thin film is an achievable setup for p-wave exciton condensation in experiments\cite{chang2013experimental, mogi2022experimental}.

	\begin{figure}[h]
		\includegraphics[width=1.0\linewidth]{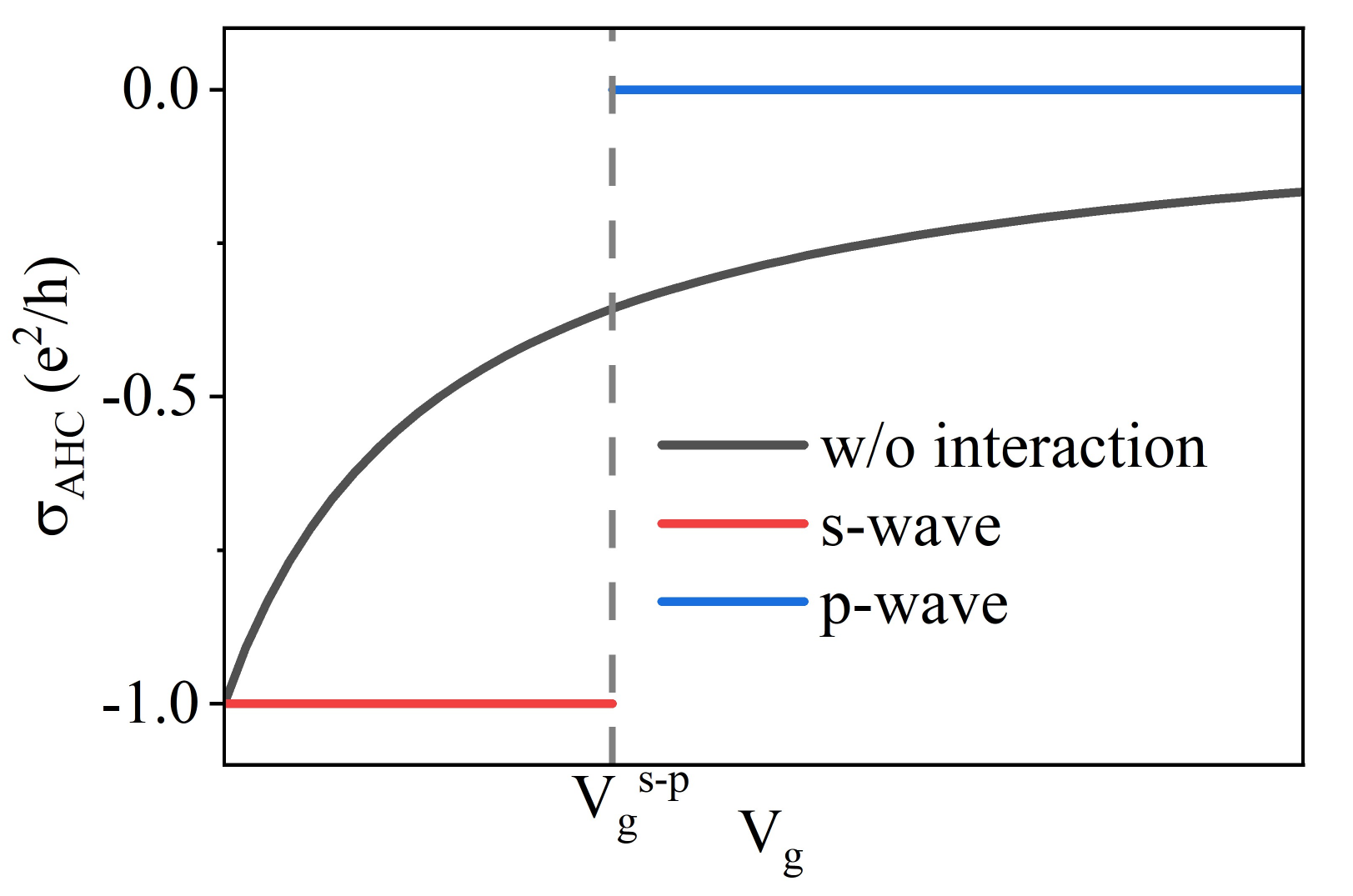}
		\caption{ \label{fig4}
			 Anomalous Hall conductivity with (red and blue lines) and without (black solid line) electron-hole interaction in the setup shown in the inset when the gating is increased.
		}
	\end{figure}
	
	\textbf{Discussion}
	
	%\textit{Expriments-}
	The change in anomalous Hall conductivity can serve as evidence for the p-wave exciton condensation in experiments\cite{RevModPhys.82.1539}, as shown in Fig.4.
	Without the interaction, the Chern numbers of the two valence bands are both $-\frac{1}{2}$ as the chiral winding numbers are both +1 in the top and bottom layers.
	Therefore, the overall anomalous Hall conductivity for this semimetalic system is $	\sigma_{\text{AHC}}=\frac{e^2}{\hbar} \int_{\text{occupied}} \frac{\mathrm{d}^2 k}{(2 \pi)^2} \Omega_{n,\boldsymbol{k}}=\left(\tilde{\Phi}_{c, \text{t}}-\tilde{\Phi}_{v, \text{b}}-1\right) \frac{e^2}{h}$.
	Including the electron-hole interaction, the system is driven into the insulating excitonic pairing phase.
	The Chern numbers of the valence bands remain $-\frac{1}{2}, -\frac{1}{2}$ if the pairing is s-wave, and become $\frac{1}{2}, -\frac{1}{2}$ if the pairing is p-wave.
	The anomalous Hall conductivity of the system becomes $\sigma^{\text{s}}_{\text{AHC}} =[(-\frac{1}{2})+(-\frac{1}{2})] \frac{e^2}{h}=-\frac{e^2}{h}$ for the s-wave pairing and $\sigma^{\text{p}}_{\text{AHC}} =[(\frac{1}{2})+(-\frac{1}{2})] \frac{e^2}{h}=0$ for the p-wave pairing.
	%	When the chiral winding numbers are both +1 in the top and bottom layers, the overall anomalous Hall conductivity for the semimetalic system without interaction is $	\sigma_{\text{AHC}}=\frac{e^2}{\hbar} \int_{\text{occupied}} \frac{\mathrm{d}^2 k}{(2 \pi)^2} \Omega_{n,\boldsymbol{k}}=\left(\tilde{\Phi}_{c, \text{t}}-\tilde{\Phi}_{v, \text{b}}-1\right) \frac{e^2}{h}$.
	%	Including the electron-hole interaction, the system is driven into the insulating excitonic pairing phase.
	%	The Chern number of the system is $-1$ if the pairing is s-wave, and becomes $-1+1=0$ if the pairing is p-wave.
	%	Therefore, the anomalous Hall conductivity of the system becomes $\sigma^{\text{s}}_{\text{AHC}} =- \frac{e^2}{h}$ for the s-wave pairing and $\sigma^{\text{p}}_{\text{AHC}} =0$ for the p-wave pairing.
	%
	This can be used as a signature for the exciton condensation not only for the magnetic topological insulator thin film, but also for the staggered graphene-based heterostructure, as the electron and hole layers are spin/valley polarized when the gating is not too high\cite{PhysRevB.107.L201119, PhysRevB.108.L041101, PhysRevB.107.L121405, PhysRevB.107.075108}.
	%	
	
	%\textit{Superfluidity-}To estimate the superfluid transition temperature, we use the BKT theory\cite{berezinskii1971destruction, KT}:  $k_{\text{B}}T_{\text{BKT}} = \frac{\pi}{8}D_s(T_{\text{BKT}})$, where $D_s$ is the superfluid weight \cite{peotta2015superfluidity, PhysRevB.95.024515}:
	%\begin{equation}
		%\begin{aligned} D^{\alpha \beta}_s= & \frac{1}{\Omega} \sum_{\boldsymbol{k}, n, m} \frac{f_F\left(E_{\boldsymbol{k},m}\right)-f_F\left(E_{\boldsymbol{k},n}\right)}{E_{\boldsymbol{k},n}-E_{\boldsymbol{k},m}}\\ 
			%&[\hat{v}^{\alpha}_{nm}(\boldsymbol{k})\hat{v}^{\beta}_{mn}(\boldsymbol{k}) -\hat{w}^{\alpha}_{nm}(\boldsymbol{k})\hat{w}^{\beta}_{mn}(\boldsymbol{k})].
		%\end{aligned}
	%\end{equation}
	%	Here $\hat{v}^{\alpha}(\boldsymbol{k})=\partial{H_{\text{MF}}}/\partial{k_{\alpha}}$, $\hat{w}^{\alpha}(\boldsymbol{k})=\gamma_z\partial{H_{\text{MF}}}/\partial{k_{\alpha}}$ and $\gamma_z$ is the Pauli matrix in the layer space.
	%
	
	%
	In summary, we discussed the high-angular momentum exciton condensation induced by band geometric effects.
	This is in analogy to the superconductivity case\cite{shi2020attractive, PhysRevLett.101.160401, PhysRevLett.106.157003, qin2019chiral}, while the attractive interaction is inherent in the exciton case.
	We also proposed experimental setups for the realization and identification for the p-wave exciton condensation.
	The high-angular momentum exciton condensation may have some exotic properties, such as fractional Josephson effects\cite{kwon2004fractional} and exciton anomalous Hall effects\cite{PhysRevB.103.165119}.
	As such, we expect which deserve further experimental and theoretical investigations.

	\textbf{Methods}

	In the heterostructure as Fig. 1(a), the interaction in the two-band basis reads:
	\begin{equation}
			\begin{aligned}
		\hat{H}_{\text{int}}= &\frac{1}{2} \sum_{i, j \in \{c, v\}} \sum_{\boldsymbol{k}_1, \boldsymbol{k}_2, \boldsymbol{q}} V_{ij,\text{int}}(q) \Lambda_{ii}\left(\boldsymbol{k}_1, \boldsymbol{k}_1-\boldsymbol{q}\right) \times \\
		 &\Lambda_{jj}\left(\boldsymbol{k}_2, \boldsymbol{k}_2+\boldsymbol{q}\right)
		c_{i\boldsymbol{k}_1}^{\dagger} c_{j\boldsymbol{k}_2}^{\dagger} c_{j \boldsymbol{k}_2+\boldsymbol{q}} c_{i \boldsymbol{k}_1-\boldsymbol{q}},
	\end{aligned}
	\end{equation}
	where
	\begin{equation}
		V_{ij,\text{int}}(q)= \begin{cases} \frac{2 \pi }{S \varepsilon_r q}, & i=j \\  \frac{2 \pi }{S \varepsilon_r q} e^{-qd}, & i \neq j\end{cases}
	\end{equation}
	is the intra- and interlayer screened Coulomb interaction.
	Now we can define $W_{ij}(\boldsymbol{k}, \boldsymbol{k^{\prime}}) =  V_{ij,\text{int}}(\boldsymbol{k}-\boldsymbol{k^{\prime}}) \Lambda_{ii}\left(\boldsymbol{k}, \boldsymbol{k^{\prime}}\right) \Lambda_{jj}\left(\boldsymbol{k^{\prime}}, \boldsymbol{k}\right)$
	as the projected interaction matrix.	
	At 0K, the Hartree-Fock mean-field Hamiltonian for this two-band system reads:
	\begin{equation}
		\hat{H}_{\text{MF}}=\sum_{\boldsymbol{k}} \Psi_{\boldsymbol{k}}^{\dagger} (H_0+H_{\text{Hartree}}+H_{\text{Fock}})  \Psi_{\boldsymbol{k}},
	\end{equation}
	where $H_{\text{Hartree}}=\left(\begin{array}{cc}
		2\pi e^2 n_{ex}d/\epsilon_r & 0 \\
		0 & -2\pi e^2 n_{ex}d/\epsilon_r
	\end{array}\right)$, and $H_{\text{Fock}}= \left(\begin{array}{cc}
		\Delta_{cc}(\boldsymbol{k}) & \Delta_{cv}(\boldsymbol{k}) \\
		\Delta_{vc}(\boldsymbol{k}) & \Delta_{vv}(\boldsymbol{k})
	\end{array}\right)$.
	\begin{equation}
		\Delta_{ij}(\boldsymbol{k})=- \sum_{\boldsymbol{k^{\prime}}} W_{ij}(\boldsymbol{k}, \boldsymbol{k^{\prime}}) \rho_{ij}(\boldsymbol{k^{\prime}}) ,
	\end{equation}
	and the density matrix
	\begin{equation}
		\rho_{ij}(\boldsymbol{k}) = \left\langle c_{j \boldsymbol{k}}^{\dagger} c_{i \boldsymbol{k}}\right\rangle - \delta_{ij} \delta_{i=v}
	\end{equation}
	is defined relative to the the fully filled valence band to aviod the double counting of the interaction.
	The exciton density is $n_{ex}=\frac{1}{S}\sum_{\boldsymbol{k}} \rho_{cc}(\boldsymbol{k})$, which appears in the Hartree term.
	Gathering all these terms together, we obtain the full Hatree-Fock Hamiltonian:
	\begin{equation}
		H_{\text{MF}}= \left(\begin{array}{cc}
			\xi_{\boldsymbol{k}} & \Delta_{cv}(\boldsymbol{k}) \\
			\Delta_{vc}(\boldsymbol{k}) & -\xi_{\boldsymbol{k}} 
		\end{array}\right) ,
	\end{equation}
	where 
	\begin{equation}
	    \begin{aligned}
		\xi_{\boldsymbol{k}} =&(\varepsilon_{c\boldsymbol{k}, \text{t}} - \varepsilon_{v\boldsymbol{k}, \text{b}} - 2V_\text{g})/2 + 2\pi e^2 n_{ex}d/\epsilon_r \\ &-\frac{1}{2}\sum_{\boldsymbol{k^{\prime}}} [W_{cc}(\boldsymbol{k}, \boldsymbol{k^{\prime}}) + W_{vv}(\boldsymbol{k}, \boldsymbol{k^{\prime}})] \rho_{cc}(\boldsymbol{k^{\prime}}).
	\end{aligned}
	\end{equation}
	The density matrix now can be written as:
	\begin{equation}
		\begin{gathered}
			\rho_{c v, \boldsymbol{k}}=-\frac{1}{2} \frac{\Delta_{cv}(\boldsymbol{k})}{\sqrt{\xi_{\boldsymbol{k}}^2+\left|\Delta_{cv}(\boldsymbol{k})\right|^2}}, \\
			\rho_{c c, \boldsymbol{k}}=\frac{1}{2}\left[1-\frac{\xi_{\boldsymbol{k}}}{\sqrt{\xi_{\boldsymbol{k}}^2+\left|\Delta_{cv}(\boldsymbol{k})\right|^2}}\right], \\
			\rho_{v v, \boldsymbol{k}}=-\rho_{c c, \boldsymbol{k}},
		\end{gathered}
	\end{equation}
	and the gap equation for the interlayer exciton condensation reads:
	\begin{equation}
		\Delta_{cv}(\boldsymbol{k})= \Sigma_{\boldsymbol{k^{\prime}}} V_{\text{inter}}(\boldsymbol{k}-\boldsymbol{k^{\prime}}) \Lambda_{cc} \left(\boldsymbol{k}, \boldsymbol{k^{\prime}}\right) \Lambda_{vv} \left(\boldsymbol{k^{\prime}}, \boldsymbol{k}\right)  \frac{\Delta_{cv}(\boldsymbol{k^{\prime}})}{2E_{\boldsymbol{k^{\prime}}}},
	\end{equation}
	where $2E_{\boldsymbol{k}}=\sqrt{\xi_{\boldsymbol{k}}^2+\left|\Delta_{cv}(\boldsymbol{k})\right|^2}$ is the the renormalized energy gap.
	We choose different initial $\Delta_{cv}$s and $\rho_{cc}$s, and solve the Hartree-Fock equations above by iteriations.
	Then the total energy density can be calculated by:
	\begin{equation}
		E_{\text{total}}=\frac{1}{2S}\sum_{\boldsymbol{k}} \text{Tr} \{\rho(\boldsymbol{k}) [H_0(\boldsymbol{k}) + H_{\text{MF}}(\boldsymbol{k})]\}.
	\end{equation}
	Then we compare the energies of different states to determine the ground state. 
	
	\textbf{Data availability}
	
	All data generated or analysed during this study are included in this published article and its supplementary information files.	
	
	\textbf{Code availability}
	
	The custom codes used in this study are available on reasonable request.
	
	%\bibliography{ref}
	

	\textbf{Acknowledgments:}
	We are supported by the National Basic Research Programs of China under Grand Nos. 2022YFA1403500 and 2021YFA1400503, the National Science Foundation of China under Grant Nos. 12234001, 11934003, and 62321004,  the Beijing Natural Science Foundation under Grant No. Z200004, the Strategic Priority Research Program of the Chinese Academy of Sciences Grant No. XDB33010400.
	The computational resources were provided by the supercomputer center in Peking University, China.

	\textbf{Author contributions}
	
	H. Y., Y. S., X. D. and X.-Z. Li designed the project; H. Y. and Y. S. performed the research and analyzed the results; H. Y., Y. S., X. D. and X.-Z. Li wrote the paper; X.-Z. Li and X. D. supervised the project.
	
	\textbf{Competing interests}
	
	The authors declare no competing interests.
	
	\textbf{Additional information}
	
	\textbf{Supplementary Information} is available for this paper.
	
	\textbf{Correspondence} and requests for materials should be addressed to Yuelin Shao, Xi Dai and Xin-Zheng Li.
	
\end{document}